\documentclass[12pt]{iopart}

\usepackage{iopams}

\usepackage[T2A]{fontenc}
\usepackage[english]{babel}
\usepackage{hyperref}
\hypersetup{
   colorlinks=true,
   linkcolor=black,
   citecolor=black,
 citebordercolor={0 0 0},
 linkbordercolor={0 0 0},
   filecolor=black,
   urlcolor=blue,
}

\def\omg{\omega}             \def\Omg{\Omega}
\def\alf{\alpha}                     
\def\gam{\gamma}             \def\Gam{\Gamma}
\def\Dlt{\Delta}                    \def\dlt{\delta}

             \def\Lam{\Lambda}
\def\vphi{\varphi}
\def\sig{\sigma}
\def\veps{\varepsilon}       \def\eps{\epsilon}
\def\dg{\dagger}
\def\ra{\rangle}                   \def\la{\langle}
\def\fns{\footnotesize}
\def\pd{\partial}
\def\dstyle{\displaystyle}
              \def\tld{\tilde}
\def\lb{\label}

\def\beq{\begin{equation}}     \def\eeq{\end{equation}}
\def\bea{\begin{eqnarray}}     \def\eea{\end{eqnarray}}
\def\bt{\begin{tabular}}            \def\et{\end{tabular}}
\def\fnt{\phantom{}}
\def\fns{\footnotesize}               \def\bs{\boldsymbol}

\begin{document}

\title[Integrals of motion and Robertson-Schr\"odinger correlated states of e.-m. field ..]
{Integrals of motion and Robertson-Schr\"odinger correlated states of electromagnetic 
field in time-dependent linear media}
\author{A.K. Angelow $^\ast$ and D.A.  Trifonov $^\dagger$}
\address{$^\ast$ Institute of Solid State Physics, Bulgarian Academy of Sciences,\\
 72 Trackia Blvd., 1184 Sofia, Bulgaria\\
  $^\dagger$ Institute of Nuclear Research, Bulgarian Academy of Sciences,\\
 Tzarigradsko chaussee 72, 1184 Sofia, Bulgaria}

\begin{abstract}
Integrals of motion and statistical properties of quantized electromagnetic field 
(e.-m. field) in time-dependent linear dielectric and conductive media are considered, 
using Choi-Yeon quantization, based on Caldirola-Kanai type Hamiltonian. 
Eigenstates of quadratic and linear invariants are constructed, the solutions 
being expressed in terms of a complex parametric function that obeys classical 
oscillator equation with time-varying frequency.   
The time evolutions of initial Glauber coherent states and Fock states are considered. 
The medium conductivity and the time-dependent electric permeability are shown to 
generate squeezing and non-vanishing covariances. In the time-evolved coherent and 
squeezed states all the second statistical moments of the electric and magnetic field 
components are calculated and shown to mminimize the Robertson-Schr\"odinger 
uncertainty relation.
\end{abstract}

\section{Introduction} 

Although the remarkable achievements have been made in classical and quantum optics,
yet probably many properties of light still remain to be uncovered.
For deeper understanding of its nature and more precise manipulation
the light needs be quantized. The method of quantizing light  propagating in free space is
well known and can be found in most quantum optics text books \cite{louisell} -- \cite{walls}.
Practically the same is the light quantization procedure in a stationary homogeneous
and isotropic dielectric media, the only new feature being the reduced light velocity.
The quantization of damped light is  somewhat more subtle.
It involves the quantum description of a single (or several) field oscillator(s) interacting
with a reservoir with a large (infinite) number of degrees of freedom (see e.g. \cite{huttner92}
and references therein) that makes the calculations rather lengthy. Besides, the
Hamiltonians they used in the development of the theory are somewhat assumed ones,
rather than having been derived in a consistent way from the classical electrodynamics.
A scheme was proposed for quantizing the damped light in conducting
(and nonstationary) linear media in publications \cite{choi03} -- \cite{choi10b}, without link to 
a reservoir and resorting to the Caldirola-Kanai Hamiltonian \cite{CK} and to the Lewis-Riesenfeld 
dynamical invariant theory \cite{lewis}. There are some reason \cite{choi10, popp} to consider the 
damped light as bio-photons in squeezed states.

The goal here is to construct Schr\"odinger Minimum Uncertainty States (SMUS -- widely known also 
as Robertson-Schr\"odinger correlated intelligent states) \cite{dodo80, trifo94} of electromagnetic
field (e.-m. field) in time-dependent dielectric and conductive linear media and to examine some of 
their properties.
We also consider the time evolution in such media of initial Glauber coherent states (CS) and Fock  
states and show that the conductivity and/or the electric permeability act as a correlating and 
squeezing factors.
We resort to the linear dynamical invariant method, developed in papers by Malkin, Man'ko
and Trifonov \cite{MMT69, MMT73} (see also \cite{holz}).

\section{Quantization of e.-m. field in time-dependent linear media} 

As in papers \cite{choi05, choi06, choi10} we consider  e.-m. field in nonstationary (isotropic and
dispersionless) linear media with no charge sources (the case of stationary conductive linear
media was considered earlier in \cite{choi03, choi04}),
\bea  
\bs{B}(\bs r,t) = \mu(t)\bs{H}(\bs r,t),\,\,
\bs{D}(\bs r,t) = \veps(t)\bs{E}(\bs r,t), \quad
\bs{j} = \sig(t)\bs{E},  \lb{B,D,j}  %
\eea
\bea  
{\rm div}\bs D =0,\quad {\rm div}\bs B =0,  \lb{divs}\\
{\rm rot} \bs H  = \frac{\pd}{\pd t}\bs{D} + \sig(t)\bs{E}, \lb{rot H}\\
{\rm rot} \bs E  = -\frac{\pd}{\pd t}\bs{B}.  \lb{rot E}
\eea
We fix the Coulomb gauge with vanishing scalar potential,
\beq\lb{ABE}
\bs{B} = {\rm rot} \bs A, \quad \bs{E} = -\frac{\pd \bs A}{\pd t},
\eeq
and write the equation for $\bs A(\bs r,t)$ (note that the term $\dot{\veps}$ is omitted in papers 
\cite{choi05, choi06})
\beq \lb{A eq}
\nabla^2 \bs{A} - \mu(\sig +\dot\veps) \frac{\pd \bs{A}}{\pd t} -
\eps\mu \frac{\pd^2 \bs{A}}{\pd t ^2} = 0.
\eeq
As usual (see e.g.  the  books  \cite{louisell, scully, walls}) we expand vector potential
$\bs{A}(\bs r,t)$ in terms of mode functions $\bs{u}_l(\bs r) = \bs e_{l,\xi}u_{l,\xi}(\bs r)$
\beq \lb{A u q}
\bs{A}(\bs r,t) = \sum_{l,\xi} \bs e_{l,\xi}u_{l,\xi}(\bs r)  q_{l}(t) ,
\eeq
where, for further convenience,  we introduced the polarization vectors $\bs e_{l,\xi}$.
The mode functions  $u_{l,\xi}(\bs r)$ are supposed to be eigenfunctions of the Laplace
operator \cite{louisell,scully,walls} ($\omg_{0,l}$ are constants of dimension of frequency),
\beq\lb{u}
\left(\nabla^2  + \frac{\omg_{0,l}^2}{c^2}\right) u_{l,\xi}(\bs r) =0,
\eeq
 to form a complete orthonormal set and  $\bs u_l$ to satisfy the transversality condition,
$\nabla\cdot \bs{u}_l  = 0 $.  {\fns For example, in case of e.-m. field in a cavity of the form of
rectangular cube of side $L=V^{1/3}$ the polarization mode in $x_1$-direction the appropriate
mode  is $\bs u_l (\bs r) =  \bs e_x \sin (l_y\pi y/L) \sin (l_z\pi z/L)$,  where $l_y,\,l_z$ are
integers.
In the case of  periodic boundary conditions, the mode functions may be taken as
\cite{louisell, walls}
\beq\lb{u_l}
 \bs u_{l,\xi}^{(\pm)}(\bs r) =V^{-1/2}\bs e_{l,\xi} \exp(\pm i\bs k_l\cdot \bs r),
\eeq
where $\bs k_l$ is the wave vector, $\bs k_l\cdot \bs e_{l,\xi} =0$. In three dimensions
$\bs k_l = 2\pi \sum_il_i\bs e_i$, $i=1,2,3$  where $l_i $ are integers and $\bs e_i$ are
orthogonal unit vectors. The set $(l_1,l_2,l_3,\xi)$ gives a mode of a given polarization.   }

From Maxwell equations (\ref{divs}) - (\ref{rot E}) it follows that (in case of linear  media
(\ref{B,D,j})) the time-dependent factors $q_l$ are to  obey the  following linear equation
(furthermore, unless otherwise stated,  we suppress the polarization index $\xi$)
\beq \lb{q eq}
 \frac{\pd^2 q_l}{\pd t ^2}  + \Gam(t) \frac{\pd q_l}{\pd t}+ \omg^2_l(t) q_l   = 0,
\eeq
where (the term $\dot{\veps}$ being omitted in \cite{choi05, choi06} )
\beq \lb{omg,Gam}
\omg_l^2(t) = \frac{\omg_{0,l}^2}{c^2\veps\mu},\quad \Gam(t) = \frac{\sig +\dot\veps}{\veps}.
\eeq
The frequencies $\omg_l$ and the damping parameter $\Gam$ are time-dependent
if the medium parameters are. Noting that the eq. (\ref{q eq}) could be obtained from the
Hamilton function
\bea\lb{Hf}
H_{k,{\rm f}}  =  \frac{1}{2}\left[ \frac{1}{\veps_0}e^{-\Lam(t)}p_l^2
+ \veps_0\omg_l^2(t) e^{\Lam(t)}q_l^2\right], \\
 \Lam(t)=\int_0^t \Gam(t')dt',
\eea
the authors \cite{choi05, choi06}  perform the quantization introducing the Hamiltonian operator
\bea\lb{H}
{\hat H} = \sum_l\left[ \frac{1}{2\veps_0}e^{-\Lam(t)}\hat{p}_l^2
+ \frac{\veps_0\omg_l^2(t)}{2} e^{\Lam(t)}\hat{q}_l^2\right] \equiv \sum_l \hat{H}_l,
\eea
were $\veps_0$ is vacuum dielectric permeability and  $\hat q_l$, $\hat p_{l'}$ are Hermitian operators satisfying canonical commutation relations,
\beq\lb{com rel}
[\hat q_l,\hat p_{l'}]= i\hbar\dlt_{l l'}, \quad
[\hat q_l,\hat q_{l'}]=[\hat p_l,\hat p_{l'}]=0.
\eeq
For constant $\Lam$ and $\omg_l$ the one mode  Hamiltonian $\hat{H}_{k}$ (and the Hamilton
function (\ref{Hf})) can be recognized as the reknown Caldirola-Kanai Hamiltonian \cite{CK},
considered for description of damped  mass oscillator motion. The vector potential operator reads
\beq \lb{hat A}
\hat{\bs{ A}}(\bs r,t) = \sum_l \bs u_l(\bs r) \hat{q}_l.
 \eeq
Replacing it in (\ref{ABE}) one obtains the quantized electric and magnetic fields
\beq \lb{hat  E}
\hspace*{-14mm}\hat{\bs{E}}(\bs{r},t) = -\frac{1}{\eps_0}
e^{-\Lam(t)} \sum_l \bs{u}_l(\bs r) \hat{p}_l ,
\eeq
\beq \lb{hat B}
\hspace*{-14mm} \hat{\bs{B}}(\bs r,t) =
e^{-\Lam(t)} \sum_l  \nabla \times\bs{u}_l(\bs r) \hat{q}_l.
\eeq
Using the time derivatives of operators $\hat{q}_l,\, \hat{p}_l$ of the form
\beq
\frac{d\hat{o}_l}{dt} = -\frac{i}{\hbar}[\hat{o}_l,\hat{H}],
\eeq
we check that all Maxwell equations (\ref{divs})-(\ref{rot E}) are satisfied by operator
fields $\hat{\bs E},\,\, \hat{\bs D}=\veps\hat{\bs E},\,\, \hat{\bs H}$ and $\hat{\bs{B}} =
\mu\hat{\bs H}$.

It worths noting here that the Hamiltonian (\ref{H}) (the Hamilton function (\ref{Hf})) that
governs the time evolution {\it does not coincide} to the field energy operator $\hat{W}$
(the field energy $W$), the latter being given by the expression
\beq\lb{W}
\hat{W} = \frac{1}{2}\sum_l \veps(t)\left(\frac{e^{-2\Lam(t)}}{\veps_0^2}\hat{p}_l^2 +
\omg_l^2(t)\hat{q}_l^2\right)  \equiv  \sum_l \hat W_l.
\eeq
This difference is a feature of dynamics of open systems. For the case of classical damped harmonic
oscillator with constant $\Gam$ and $\omg$ it was first noted by Bateman in 1931 \cite{bateman}.
The stationary damped oscillator was later quantized by Caldirola and Kanai \cite{CK} using
Bateman classical Lagrangian \cite{bateman}. Now we see that quantized e.-m. field in conducting media
could be regarded as a set of (noninteracting) damped oscillators. \\
The Hamiltonian operator (\ref{H}) has the form of a sum of oscillators with varing mass 
$m(t) = \varepsilon_0 e^{-\Lambda(t)}$ and varing frequency $\omega_l(t)$ (according to eq. (\ref{Hf}) $\omega_l(t) = \omega_{0l}/c\sqrt{\mu\varepsilon}$). 
For e.-m. field in free space $\mu\veps = 1/c^2$, $\omega_l =\omega_{l0}$, $\Lambda$ is vanishing 
and $H_l$ equals the stationary oscillator Hamiltonian. This gives us a confidence to consider the canonical operators $q_l$ and $p_l$ as the quadratures of photon creation and annihilation operators.

\section{Integrals of motion -- connection with classical Ermakov equations}  

Integrals of motion $\hat I$ of quantum system with Hamiltonian $\hat H$ are defined as
solutions to the equation
\beq\lb{I}
\frac{\pd \hat I}{\pd t} -\frac{i}{\hbar}[\hat I,\hat H] = 0.
\eeq
The canonical commutation relations (\ref{com rel}) show that quadratic in $\hat q$ and
$\hat p$ Hamiltonians admit linear in $\hat q$ and $\hat p$ dynamical invariants. In refs.
\cite{MMT73} a family of (non-Hermitian) invariants $\hat A$ for the general nonstationary
quadratic Hamiltonian
\beq\lb{H2}
\hat H = \frac{1}{2}\left[ a(t)\hat{p}^2 +b(t)(\hat{p}\hat{q}+\hat{q}\hat{p}) + c(t)
\hat{q}^2\right]
\eeq
have been constructed in the form
\beq\lb{A(t)}
\hat A(t) = \sqrt{\frac{a}{2\hbar}}\left[\eps\hat p +\frac 1a\left(\eps b-\dot\eps-\frac{\dot a}{2a}
\eps\right)\hat q \right] ,
\eeq
where $\eps$ is any solution of the second order equation (classical oscillator equation)
\bea
\ddot\eps + \Omg^2(t)\eps =0, \lb{eps}\\
\Omg^2 = ac + b\frac{\dot a}{a} +\frac{\ddot a}{2a} - \frac{3\dot a^2}{4a^2} -b^2-\dot b.
\lb{Omg}
\eea
The commutator $[\hat A,\hat A^\dg]$ reads
\beq
[\hat A,\hat A^\dg] = \frac i2(\eps\dot\eps^* - \eps^*\dot\eps) \equiv \frac i2 w.
\eeq
This shows that if the Wronskian $w$ of the equation (\ref{eps}) is fixed to $-2i$  (as in
\cite{MMT69, MMT73}), then the invariants $\hat A$ and $\hat A^\dg$ are boson ladder operators.
(It is worth noting that  real solutions $\eps(t)$ are also admissible, they correspond to Hermitian 
invariants).
The relation $w=-2i$ is identically satisfied with $\eps$ of the form 
$\eps = |\eps|\exp(i\int dt'/|\eps(t')|^2)$.
Now eq. (\ref{eps}) leads to the Ermakov equation \cite{ermakov} for $|\eps(t)|$,
\beq\lb{Erm eq}
\frac{d^2|\eps|}{dt^2} + \Omg^2(t)|\eps| - \frac{1}{|\eps|^3} =0.
\eeq
At
\beq\lb{at}
a(t) = \veps_0^{-1} e^{-\Lam(t)},\quad b(t) =0, \quad c(t) = \veps_0\omg^2(t) e^{\Lam(t)}
\eeq
the Hamiltonian (\ref{H2}) recovers the nonstationary one-mode Caldirola-Kanai Hamiltonian
$\hat H_l$, eq. (\ref{H}). Therefore at (\ref{at}) the ladder operator linear invariants
$\hat A(t)$, eq. (\ref{A(t)}), are dynamical invariants for the damped oscillator. \\
Thus for each mode $l$ of the quantized e.-m. field we have dynamical invariants in the form 
of boson annihilation operators
\beq\lb{A_l}
\hat A_l =  \frac{1}{\sqrt{2\hbar\veps_0}}e^{-\frac 12\Lam(t)}\left[\eps_l \hat p_l -
\veps_0e^{\Lam(t)}\left(\dot\eps_l - \frac{1}{2}\dot\Lam \eps_l\right)\hat q_l \right]\, ,
\eeq
where $\eps_l$ satisfy the equations
\bea
\ddot\eps_l + \Omg_l^2(t)\eps_l =0,   \lb{eps_l}  \\
\Omg_l^2(t) = \omg_l^2(t) - \frac 12 \ddot\Lam - \frac 14\dot\Lam^2,   \lb{Omg_l}
\eea
and $\eps_l\dot\eps_l^* - \eps_l^*\dot\eps_l = -2i$. Note that under the latter conditions 
$[\hat A_l, \hat A_l^\dg] =1$ and all $|\eps_l|$ are solutions to the Ermakov equaton (\ref{Erm eq}).   

For general quadratic Hamiltonian (\ref{H2}) the linear invariants $\hat A(t)$ and their Hermitian
combination $\hat A^\dg(t)\hat A(t)$ have been diagonalized in \cite{MMT73}:
\beq\lb{|alf t>}
\hat A(t) |\alf;t\ra = \alf|\alf;t\ra ,\quad  \hat A^\dg(t)\hat A(t) |n;t\ra = n|n;t\ra,
\eeq
where $\alf\in C$ and $n=0,1,2, \ldots$.
For our quadratic Hamiltonians $\hat H_l$ and the invariants $\hat A_l$, $\hat A_l^\dg \hat A_l$
formulas (44) and (46) of the first paper in \cite{MMT73} produce the eigenfunctions
$\psi_{\alf_l}(q_l,t) = \la q_l|\alf_l;t\ra$,\, $\psi_{n_l}(q_l,t) = \la q_l|n_l;t\ra$  \,\,
($ a = \veps_0^{-1} e^{-\Lam(t)} = a(t)$),
\beq\lb{psi_alf t}
\psi_{\alf_l}(q_l,t) = \psi_0(q_l,t)\exp\left[ \sqrt{\frac{2}{a \hbar}}\,\frac{\alf_l}{\eps_l}
q_l - \frac{\eps_l^*}{2\eps_l}\alf_l^2 - \frac{1}{2}|\alf_l|^2 \right] \, ,
\eeq
\beq\lb{psi_n t}
\psi_{n_l}(q_l,t) = \psi_0(q_l,t) \frac{(\eps_l^*/2\eps_l)^{n_l/2}}{\sqrt{n_l!}} H_{n_l}(x_l),
\quad x_l = \frac{q_l}{|\eps_l|\sqrt{a}}\, ,
\eeq
where $H_n(x)$ are Hermite polynomials, $\psi_0(q_l,t)$ are the ground state wave functions,
\beq 
\psi_0(q_l,t)  =  \left(\eps_l(\pi a \hbar)^{\frac 12}\right)^{-\frac 12}\exp\left[
\frac{i}{2a\hbar} \left(\frac{\dot\eps_l}{\eps_l}+\frac{\dot a}{2a}\right)q_l^2 \right] \, .
\eeq
These time-dependent wave functions are normalized solutions to the Schr\"odinger equation
$i\hbar \pd_t \psi = \hat H_l\psi$ with Hamiltonian $\hat H_l$ given by eq. (\ref{H}). 
Since $\hat A(t)$ and $\hat A_l^\dg(t)\hat A_l(t)$ are dynamical invariant, the eigenvalues 
$\alf_l$ and $n_l$ are constant in time.

The system of $|\alf_l;t\ra$ is overcomplete  in the one mode Hilbert space ${\cal H}_l$
(the set of $|n_l;t\ra$  being complete):
\beq
\frac{1}{\pi}\int |\alf_l;t\ra\la t;\alf_l| d^2\alf_l = \sum_{n_l} |n_l;t\ra\la t; n_l| = \mathbf{ 1}_l.
\eeq
According to the terminology of Refs. \cite{MMT69, MMT73} the states $|\alf_l;t\ra$ may be called CS 
of nonstationary system with Hamiltonian $\hat H_l$, eq. (\ref{H}). Since the e.-m. field Hamiltonian $\hat H$,
eq. (\ref{H}), is a sum over $l$, CS for e.-m. field with finite number of modes are product over $l$ of
one mode CS $|\alf_l;t\ra$.

\vspace{3mm}

{\fns {\bf Remark 1.} To make connection with the Choi and Yeon treatment  \cite{choi05, choi06} we put
\bea
\eps_l(t) = \frac{1}{\sqrt{\dot{\gam}_l}}e^{i\gam_l(t)},   \lb{eps - gam} \\
\dot\eps_l = \frac{1}{2\sqrt{\dot{\gam}_l}}\left[\dot\Lam + \frac{2\dot M_l}{M_l} +i\dot\gam_l \right]
e^{i\gam_l(t)}. \lb{deps - M}
\eea
Then we find that the two real parametric functions $\gam_l(t)$ and $M_l(t)$ have to obey the
following system of (second order nonlinear) equations
\bea
\ddot\gam_l = -\dot\gam_l\left( \dot\Lam + 2\frac{\dot M_l}{M_l}\right), \lb{ddot gam}\\
\ddot M_l = M_l\left[\left(\dot\gam_l^2 - \omg_l^2(t)\right)  - \dot M_l\frac{\sig +\dot\veps}{\veps}\right].
\lb{ddot M}
\eea
Up to the term $\dot M_l\dot\veps/\veps$ these equations do coincide with  eqs. (17), (18) in Ref.
\cite{choi05} and eqs. (7), (8) in \cite{choi06}. Evidently this term was omitted in \cite{choi05, choi06}.
Under the substitutions (\ref{eps - gam}), (\ref{deps - M}) our linear  and quadratic invariants
$\hat A_l(t)$, $\hat A_l^\dg(t) \hat A_l(t)$ and their eigenstates $|\alf_l;t\ra$, $|n_l;t\ra$ will
coincide with Choi-Yeon operators $\exp(i\gam_l(t))\hat a_l(t)$, $\hat a_l^\dg(t)\hat a_l(t)$ and their
eigenstates $|\alf_l(t)\ra$, $|\phi_{n,l}\ra$ \cite{choi05, choi06}, if the parametric functions
$\gam_l(t)$, $M_l(t)$ in \cite{choi05, choi06} are subject to the equations (\ref{ddot gam}),
(\ref{ddot M}) and  their $\Gam(t)$ is replaced with our $\Gam(t)$, eq. (\ref{omg,Gam}) .   }

\section{Evolution of CS and photon number states in time-dependent linear media}  

Let $\hat a_{\omg_{0,l}}$ be photon annihilation operator corresponding to the constant frequency $\omg_{0,l}$.
In terms of Hermitian operators $\hat q_l$ and $\hat p_l$ its expression reads
\beq 
\hat a_{\omg_{0,l}} = \sqrt{\frac{\veps_0\omg_{0,l}}{2\hbar}}\left(\hat q_l +\frac{i}
{\veps_0\omg_{0,l}}\hat p_l\right).
\eeq
One has $[\hat a_{\omg_{0,l}}, \hat a_{\omg_{0,l}}^\dg] =1$.  Glauber coherent states (CS)
$|\alf_l\ra$ of one-mode e.-m. field are defined as eigenstates of the photon annihilation operator
$\hat a_{\omg_{0,l}}$,
\beq 
\hat a_{\omg_{0,l}}|\alf_l\ra = \alf_l|\alf_l\ra \quad \rightarrow \quad
|\alf_l\ra = \exp\left(\alf_l^*\hat a_{\omg_{0,l}}-\alf_l \hat a_{\omg_{0,l}}^\dg\right)|0\ra.
\eeq
CS of the $N$-mode field are given as product of the $l$-one-mode CS.

The $n_l$ photon states $|n_l\ra$ (the Fock states or the number states) are eigenstates of photon
number operators $\hat n_l = \hat a_{\omg_{0,l}}^\dg \hat a_{\omg_{0,l}}$,
\beq
\hat n_l |n_l\ra = n_l|n_l\ra  \quad \rightarrow \quad |n_l\ra =
\frac{(\hat a_{\omg_{0,l}})^{n_l}}{\sqrt{n_l!}} |0\ra.
\eeq
The normalized wave functions of these states are
\beq\lb{psi_alf}
\psi_{\alf_l}(q_l) = \psi_0(q_l)\exp\left[ \sqrt{\frac{2\veps_0 \omg_{0,l}}{\hbar}}\,\alf_lq_l -
\frac{1}{2}\alf_l^2 - \frac{1}{2}|\alf_l|^2 \right]\, ,
\eeq
\beq\lb{psi_n}
\psi_{n_l}(q_l) = \psi_0(q_l) \frac{2^{-n_l/2}}{\sqrt{n_l!}} H_{n_l}(q_l\sqrt{\omg_{0,l}\veps_0}),
\eeq
where $H_n(x)$ are Hermite polynomials, and $\psi_0(q_l)$ is the ground state wave function,
\beq 
\psi_0(q_l) =  \left(\frac{\veps_0\omg_{0,l}}{\pi \hbar}\right)^{\frac 14} \exp\left[
-\frac{\veps_0\omg_{0,l}}{2\hbar}\,q_l^2 \right]\, ,
\eeq

The photon statistics in Glauber CS $|\alf_l\ra$ is Poissonian with mean $\la\alf_l|\hat n_l|\alf_l\ra =
|\alf_l|^2$.  They are remarkable also with minimal fluctuations in $\hat q_l$ and $\hat p_l$:
the variances of the dimensionless quadratures of photon ladder operators 
$$\hat Q_l=\sqrt{\frac{\veps_0\omg_{0,l}}{\hbar}} \hat q_l ,\qquad \hat P_l = \sqrt{\frac{1}{\veps_0\omg_{0,l}\hbar}}\hat p_l $$ 
are equal at the lowest possible level,
\beq
(\Dlt Q_l)^2_{\alf_l} = (\Dlt P_l)^2_{\alf_l} = \frac 12,
\eeq
 minimizing the Heisenberg uncertainty relation (UR)
\beq\lb{HUR}
(\Dlt Q_l)^2  (\Dlt P_l)^2  \geq  \frac 14 \left| \la [\hat Q_l, \hat P_l]\ra\right|^2  = \frac 14.
\eeq
In CS $|\alf_l\ra$ the $\hat q_l$-$\hat p_l$ covariance ${\rm Cov}(\hat q_l,\hat p_l)_{\alf_l}$ is
vanishing.

 Glauber CS and number states are both time-stable in vacuum and in stationary nonconductive
linear media as well.  In these cases the fluctuations of electric and magnetic field in CS are also
minimal, as one can see from eqs. (\ref{hat E}), (\ref{hat B}).
In conductive and/or nonstationary linear media  Glauber CS and photon number states can be shown
to be both time-unstable. Our next aim is to examine their time evolution.

Our time-dependent states $|\alf_l;t\ra$ and $|n_l;t\ra$ are completely determined by a solution
$\eps_l(t)$ to auxiliary classical equation (\ref{eps_l}). Therefore initial conditions $\eps_l(0),\,
\dot\eps_l(0)$ of $\eps_l(t)$ determine uniquely the evolution of quantum states of e.-m. field in nonstationary
media (\ref{B,D,j}). One can check that at
\beq\lb{eps(0)}
\eps_l(0) = \frac{1}{\sqrt{\Omg_l(0)}},\quad \dot\eps_l(0) = i\sqrt{\Omg_l(0)}
\eeq
the wave functions $\psi_{\alf_l}(q_l,0\ra$ and $\psi_{n_l}(q_l,0\ra$ recover the Glauber CS
$\psi_{\alf_l}(q_l\ra$, eq. (\ref{psi_alf}), and Fock states $\psi_{n_l}(q_l\ra$, eq. (\ref{psi_n}) with
$\omg_{0,l}$ replaced by $\Omg_l(0)$. If in addition $\dot\Lam(0) = 0$ then $\Omg_l(0) = \omg_{l0}$.) 
This means that the initial CS $|\alf_l\ra$ and the initial photon number states $|n_l\ra$ at $t>0$ evolve (in linear media (\ref{B,D,j}))  into states $|\alf_l;t\ra$ and $|n_l;t\ra$.
At $t>0$ the states $|\alf_l;t\ra$ and $|n_l;t\ra$ deviate from the form of Glauber CS and photon number
states, that is they are no more eigenstates of the Schr\"odinger operators $\hat a_{\omg_{0,l}}$ and
$\hat a_{\omg_{0,l}}^\dg \hat a_{\omg_{0,l}}$ correspondingly. They remain eigenstates of the dynamical
invariants $\hat A_l(t)$ and $\hat A_l^\dg(t)\hat A_l(t)$ with thae same constant eigenvalues $\alf_l$ and $n_l$. As a result the photon statistics in
$|\alf_l;t\ra$ is no more Poissonian, the fluctuations of $\hat Q_l$ and $\hat P_l$ deviate from their
minimal value of $1/2$ and do not minimize anymore the Heisenberg UR. It is worth noting at this point
that the invariant $\hat A_l(t)$ is a non-Hermitian linear combination of the photon creation and
annihilation operators $\hat a_{\omg_{0,l}}$ and $\hat a_{\omg_{0,l}}^\dg$.

The properties of states with Gaussian wave functions like $\psi_{\alf_l}(q_l,t)$ (eigenfunctions
of non-Hermitian linear combinations of photon creation and annihilation operators) are
well examined in the literature under the names: CS of nonstationary systems \cite{MMT69, MMT73},
Stoler states \cite{stoler}, squeezed states (SS) \cite{holenhorst}, two-photon states
\cite{yuen}, correlated states \cite{dodo80}, generalized intelligent states \cite{trifo94}.
The name SS refers to their property of "squeezing" the fluctuations of $\hat Q$ or $\hat P$.
The name 'correlated CS' underlines the existence of nonvanishing $\hat q$-$\hat p$ covariance
and the name 'generalized intelligent' stresses on their 'generalized minimum uncertainty
property' in the sense of minimization of the more general UR of Robertson-Schr\"odinger
\cite{RS},
\beq\lb{RS UR}
(\Dlt q_l)^2  (\Dlt p_l)^2  \geq  \frac{\hbar^2}{4} + {\rm Cov}^2(q_l,p_l).
\eeq
In the time-evolved CS $|\alf_l;t\ra$ we have \cite{trif74}
\beq\lb{q,p Dlt}
\bt{ll}
$\dstyle (\Dlt q_l)^2_{\alf_l} = \frac{\hbar a}{2}\rho_l^2,\quad \rho_l = |\eps_l(t)|,$\\
$\dstyle (\Dlt p_l)^2_{\alf_l} = \frac{\hbar}{2a}\left[\frac{1}{\rho_l^2}  + \left(\dot\rho_l(t) +
\frac{\dot a}{2a}\rho_l\right)^2 \right] $
\et
\eeq
and the covariance \cite{angelow} (let us remind that now  $a = \veps_0^{-1}e^{-\Lam(t)}$)
\beq\lb{q-p cov}
{\rm Cov}(q_l,p_l)_{\alf_l} = -\frac{\hbar}{2} \rho_l
\left(\dot\rho_l + \frac{\dot a}{2a}\rho_l\right).
\eeq
It is seen that these three second moments do minimize  UR (\ref{RS UR}). In the Yuen \cite{yuen}
$(u,\,v)$-parameters $\hat A_l = \tld u_l(t) \hat a_{\omg_{0,l}}  +\tld v_l(t) \hat a_{\omg_{0,l}}^\dg$,
the variances of the dimensionless canonical operators take the form
$$ (\Dlt Q_l)^2_{\alf_l} = \frac 12|\tld u_l(t) -\tld v_l(t)|^2,\quad
(\Dlt P_l)^2_{\alf_l} = \frac 12|\tld u_l(t)+\tld v_l(t)|^2,$$
which clearly show that the fluctuations in $Q_l$ ($P_l$) can be 'squeezed' much below their ground state
value of  $1/2$ when $\tld v_l \rightarrow \tld u_l$ ($\tld v_l \rightarrow - \tld u_l$).

In the case of stationary media ( constant $\veps,\, \mu$ and $\sigma$)  the 'frequencies' $\Omg_l$
are constant, $\Omg_l^2 = \omg_l^2 - \sig^2/4\veps^2$. In this case the solutions to
eq. (\ref{eps_l}) are
\beq\lb{sol1}
\eps_l(t) = \rho_l \exp(i\Omg_l t),\quad \rho_l =  \Omg_l^{-1/2} = {\rm constant},
\eeq
and our formulas (\ref{q,p Dlt}), (\ref{q-p cov}) show that the $\hat q_l$-$\hat p_l$ covariances are
time-independent, ${\rm Cov}(q_l,p_l)_\alf = \hbar \sig/4\veps\Omg_l$, and
$$ (\Dlt q_l)^2_{\alf_l}(\Dlt p_l)^2_{\alf_l} =
\frac{\hbar^2}{4} \left(1+\left(\frac{\sig}{2\veps\Omg_l}\right)^2\right).    \eqno(53a)$$
If in addition $\sigma=0$ (i.e. $\dot\Lam=0$), then  the $\hat q_l$-$\hat p_l$ covariances are vanishing
and  the fluctuations of the dimensionless $\hat Q_l$ and $\hat P_l$ become equal to $1/\sqrt{2}$.

Thus on the initial Glauber CS of e.-m. field the medium conductivity and the time dependence of electric
and/or magnetic permeability  {\it act  as $q_l$-$p_l$ correlating and squeezing  factors}.
For such stationary conductive media note the dumping factor $\exp(-\sig t/\veps)$ in the variances
of $\hat q_l$ and the amplifying factor $\exp(\sig t/\veps)$ in the variances of $\hat p_l$.

{\fns {\bf Remark 2.}
Exact solutions of eq. (\ref{eps}) for $\eps(t)$ are known for several cases of $\Omg(t)$.
For example if  $\Omg(t)$ decreases in time as $\Omg(t) = 1/(\omg^{-1}_0 + t)$ the general solution
is
\beq\lb{eps gen1}         
\eps(t) = c_1\sqrt{\tau(t)}\cos(s_0\,{\rm ln}\tau(t)) + c_2\sqrt{\tau(t)}\sin(s_0\,{\rm ln}\tau(t)),
\eeq
where  $\tau(t) = t+1/\omg_0$, $s_0=\sqrt{3}/2$, and $c_{1,2}$ are arbitrary constants.
For  $\Omg^2 = \omg^2_2 +\omg^2_1\cos(2t) $ the eq. (\ref{eps}) is known as Mathiew equation.
Let us note that the new auxiliary functions $\eps_l^\prime(t) := \eps_l(t)\exp(-\Lam/2)$ have to obey the
classical damped oscillator equation (\ref{q eq}):
\beq\lb{eps'}
\ddot\eps_l^\prime +\dot\Lam \dot\eps_l^\prime + \omg_l^2(t) \eps^\prime = 0.
\eeq
For $\dot\Lam=\eta$, $\dot\eta=0$ and constant $\omg_l$ solution for $\eps_l^\prime(t)$ in terms of Bessel
functions $J_1$, $Y_1$ is also known \cite{pedrosa09}.  Noting the analogy of eq. (\ref{eps}) to the
one dimensional stationary Schr\"odinger equation $ (\hat p/2m + U(x))\psi(x) = E \psi(x)$  one can find
solutions $\eps(t)$ in all cases of  $U(x)$ where $\psi(x)$ is known, using the correspondence
$\Omg^2(x) = 2m(E-U(x))/\hbar$  \cite{MMT69}.}

\vspace{5mm}

\section{Statistical properties of quantized  field in media}  
For the study of  statistical properties of quantized  field in the time-evolved CS $|\alf_l;t\ra$ and number
states $|n_l;t\ra$ it is convenient to express Hermitian operators $\hat q_{l}$,  $\hat p_{l}$  in terms
of the dynamical invariants $\hat A_{l}(t)$, $\hat A_{l}^\dg(t)$.
We have
\beq\lb{qp AA+}
\bt{ll}
$\hat q_{l} = \nu_l\hat A_{l}^\dg(t) + \nu_l^*\hat A_{l}(t))$, \\[1mm]
$\hat p_{l} = \mu_l\hat A_{l}^\dg(t) + \mu_l^*\hat A_{l}(t)$,
\et
\eeq
where
\beq\lb{mu, nu}
\bt{ll}
$\nu_l = \tld\nu_l e^{-\Lam(t)/2}, \quad \tld\nu_l =  -i\left(\frac{\hbar}{2\veps_0}\right)^{1/2}  \eps_l $,\\[2mm]
$\mu_l = \tld\mu_l  e^{\Lam(t)/2},  \quad \tld\mu_l  =  -i\left(\frac{\hbar\veps_0}{2}\right)^{1/2} 
\left(\dot\eps_l -\frac 12 \dot\Lam\eps_l\right)$. 
\et
\eeq
In terms of the above (time-dependent) parameters $\mu_l$, $\nu_l$ the three second statistical moments of the photon annihilation operatorquadratures in CS $|\alf_l;t\ra$ read 
\beq
 (\Dlt q_l)^2 = |\mu_l|^2,\quad
(\Dlt p_l)^2 =  |\nu_l|^2, \quad 
{\rm Cov}(q_l,p_l) = \hbar {\rm Re}(\mu_l\nu_l^*).
\eeq
The mean number of photons of mode $l$ (that is $\la \hat{a}_l^\dg \hat{a}_l\ra$)  in $|\alf_l;t\ra$ is found as
\begin{eqnarray}
 \la \hat{a}_l^\dg \hat{a}_l\ra &=& \left(|\bar{\mu}_l|^2+|\bar{\nu}_l|^2\right)|\alf_l|^2 + {\rm Re}\left((\bar{\nu}_l^2 -\bar{\mu}_l^2){\alf_l^*}^2\right) \nonumber\\ 
&\,& + \frac 12  \left(|\bar{\mu}_l|^2 +|\bar{\nu}_l|^2  - 1 \right), 
\end{eqnarray}
where $\bar{\mu}_l = -i\mu_l /\sqrt{\veps_0\omg_{l0} \hbar}$,  \, $\bar{\nu}_l = \nu_l\,\sqrt{\veps_0\omg_{l0}/\hbar}$ and we have taken into acount that $2{\rm Re}(\bar{\mu}_l^*\bar{\nu}_l) = 1$. Under the initial conditions (\ref{eps(0)}) with $\dot{\Lambda}(0)=0$ one has $\bar{\mu}_l = 1/\sqrt{2} =\bar{\nu}_l$ and then $\la \hat{a}_l^\dg \hat{a}_l\ra = |\alf_l|^2$. In the orthonormal states $|n_l;t\ra$ the mean photon number is 
\beq 
\la t;n_l| \hat{a}_l^\dg \hat{a}_l|n_l;t\ra = \left(|\bar{\mu}_l|^2+|\bar{\nu}_l|^2\right)  n_l 
                     + \frac 12  \left(|\bar{\mu}_l|^2 +|\bar{\nu}_l|^2  - 1 \right).
\eeq
Since in any state $(\Dlt Q_l)^2 + (\Dlt P_l)^2 \geq 1$ we have the iquality 
\beq
\la t;n_l| \hat{a}_l^\dg \hat{a}_l|n_l;t\ra  \geq n_l.
\eeq
 
In order to find the field component operators and their statistical moments we shall consider the case of periodic  boundary conditions with {\it complex mode functions} (\ref{u_l}).
With these modes the Hermitian operator of the vector potential, which obeys the equation (\ref{A eq})
takes the form
\beq\lb{hat A2}
\hat {\bs A}(\bs r,t) = \sqrt{\frac{\hbar}{2\veps_0}}\exp^{-\frac 12\Lam(t)}\sum_{l,\xi}\bs e_{l,\xi}
\left[ u_{l,\xi}^+(\bs r)\eps_l \hat A_{l}(t) + h.c.\right] ,
\eeq
where $\bs e_{l,\xi}$ is the polarization vector of mode $l$.
The operators $\hat{\bs E}$ and $\hat{\bs B}$ are obtained from this $\hat{\bs A}$ via the relations (\ref{ABE}):
\beq\lb{hat EB}
\bt{ll}
$\hat{\bs E}(\bs r,t) = \sqrt{\frac{\hbar}{2\veps_0}} e^{-\frac 12\Lam(t)}\sum_{l,\xi} \bs e_{l,\xi}
\left[\left(\frac 12\dot\Lam\eps_l -\dot\eps_l\right) u^{+}_{l,\xi}(\bs r)\hat A_{l}(t) + h.c. \right], $\\[3mm]
 $\hat{\bs B}(\bs r,t) = i\sqrt{\frac{\hbar}{2\veps_0}}e^{-\frac 12\Lam(t)}\sum_{l,\xi} \bs k_l\times
\bs e_{l,\xi}\left [  u^{+}_{l,\xi}(\bs r) \eps_l \hat A_{l}(t) - h.c.\right].$
 \et
\eeq
The commutators between the $j$ and $m$ components of $\hat{\bs E}_l(\bs r,t)$ and $\hat{\bs B}_l(\bs ,t)$
are  $C$-numbers, vanishing for $j=m$:
\bea
\fnt{}\hspace{-10mm}
[\hat E_{l,j}(\bs r,t),\hat B_{l,m}(\bs r,t)] =  i\frac{\hbar}{\veps_0 V } e^{-\Lam(t)} \times \nonumber\\
\sum_\xi e_{l,\xi,j}(\bs k_l \times \bs e_{l,\xi})_j\,{\rm Re}(\dot\eps_l \eps_l^* - |\eps_l|^2 \dot\Lam/2)\, \dlt_{jm}.
\eea

{\bf One mode field.}\, We shall calculate the first and second statistical moments of electric and
magnetic field in the time-evolved CS $|\alf_l;t\ra$ in the case of one mode polarized field
propagating along $x$-direction. In terms of parametric functions $\gam_l(t)$, $M_l(t)$ (see Remark 1.)
the  first moments of $\hat E_l$ and $\hat B_l$ and the means $\la \hat E_l^2\ra_\alf$,
$\la \hat B_l^2\ra_\alf$ have been calculated by Choi \cite{choi06}.

Using the mode functions (\ref{u_l}), the expressions (\ref{hat EB}) for $\hat E_l$ and $\hat B_l$  and the eigenvalue property
$$\hat A_{l}(t)|\alf_{l};t\ra = \alf_{l}|\alf_{l};t\ra,$$
we get the first  moments of  $\hat E_l$,  $\hat B_l$ and $[\hat E_l, \hat B_l] $ (of one mode polarized field) in CS $|\alf_l;t\ra$ in the forms
\bea\lb{<E_l>}
\fnt{} \hspace{-10mm}
\la \hat E_l\ra_{\alf_l} =  \sqrt{\frac{\hbar}{2\veps_0 V}} e^{-\frac 12\Lam(t)}|\alf_l| \times \nonumber\\
\left[\left(\dot\Lam \rho_l -\dot\rho_l\right)\cos(k_l x \!+\! \vphi_l (t) \!+\!\theta_l)  - \frac{1}{\rho_l}
\sin((k_l x \!-\! \vphi_l (t) \!+\!\theta_l) \right],
\eea
\bea\lb{<B_l>}
\fnt{} \hspace{-10mm}
\la \hat B_l\ra_{\alf_l} =  -k_l\sqrt{\frac{2\hbar}{\veps_0 V}} e^{-\frac 12\Lam(t)}|\alf_l| \rho_l
\sin((k_l x  \!+\! \vphi_l (t) \!+\! \theta_l),
\eea
\beq
\la [E_l,B_l]\ra_\alf = -k_l \frac{\hbar}{2\veps_0 V} e^{-\Lam(t)} {\rm Re} (\dot\Lam |\eps_l|^2/2  -
\dot\eps_l  \eps_l ^* ).
\eeq
where $\rho_l =|\eps_l|$, $\vphi_l = \arg\eps_l = \int_0^t dt' \rho^{-2}(t')$ and $\theta_l = \arg\alf_l
= {\rm const}$.\\
The three second moments are found as
\beq\lb{m2 E}
(\Dlt E_l)^2_\alf = \frac{\hbar}{2\veps_0 V} e^{-\Lam(t)}\,|\dot\Lam |\eps_l|^2/2 -\dot\eps_l|^2,
\eeq
\beq\lb{m2 B}
(\Dlt B_l)^2_\alf = k_l^2 \frac{\hbar}{2\veps_0 V} e^{-\Lam(t)} | \eps_l|^2 ,
\eeq
\beq\lb{cov EB}
{\rm Cov}(E_l,B_l)_\alf = -k_l \frac{\hbar}{2\veps_0 V} e^{-\Lam(t)} {\rm Im} (\dot\Lam |\eps_l|^2/2 -
\dot\eps_l  \eps_l^* ).
\eeq
Note also the dumping factors $e^{-\Lam/2}$ or $e^{-\Lam}$ in the expressions of all the above averages.\\
The same exponential damping factor  $e^{-\Lam}$ occurs in the expressions of mean field energy 
$\la \hat W_l\ra $ in time-evolved CS $|\alf_l;t\ra$ and time-evolved  number states $|n_l; t\ra$. Using the expression (\ref{W}) for the energy operator and eqs. (\ref{qp AA+}),  (\ref{|alf t>}) we find 
\beq\lb{<W>_alf}
\la \hat W_l\ra_\alf = \frac{\veps(t)}{2\veps_0^2} e^{-\Lam(t)}\left[ \tld a_l^*(t) \alf^2 +\tld a_l(t){\alf_l^*}^2 
+2\tld c_l(t)|\alf_l|^2 + \tld c_l(t)\right] , 
\eeq
\beq\lb{<W>_n} 
\la \hat W_l \ra_n = \frac{\veps(t)}{\veps_0^2} e^{-\Lam(t)} \tld c_l(t)\, ( n_l + 1/2), 
\eeq
where 
\beq\lb{tld a}
\tld a_l = \tld{\mu}_l^{2}+\omg_l^2\veps_0^2 \tld{\nu}_l^{^2}, \quad \tld c_l = |\tld \mu_l|^2 + |\tld \nu_l|^2\omg_l^2\veps_0^2, 
\eeq 
$\tld \mu_l$ and $\tld \nu_l$ being expressed in terms of $\Lam$ and $\veps_l$ acording to eq. (\ref{mu, nu}).

Next we check the Robertson-Schr\"odinger UR for the one mode fields $\hat E_l,\,\hat B_l$ and find
that the obtained three second moments (\ref{m2 E}) - (\ref{cov EB}) do minimize it,
\beq
\hspace*{-10mm}
\left(\Dlt E_l\right)_\alf^2\left(\Dlt B_l\right)_\alf^2 - {\rm Cov}^2(E_l,B_l)_\alf^2 =
\frac{\hbar^2}{4}\left|\la[E_l,B_l]\ra_\alf\right|^2.
\eeq
Thus the time-evolved CS $|\alf_l;t\ra$ in nonstationary and/or conductive media are
Robertson-Schr\"odinger intelligent states with respect to the photon ladder operator quadratures
$\hat q_l$, $\hat p_l$, and with respect to the electric and magnetic field components as well.
In their time development these states can exhibit $q_l$-$p_l$ and $E_l$-$B_l$ correlations and squeezing.

\section*{Concluding Remarks}

We have constructed boson like integrls of motion $\hat A_l(t)$ of electromagnetic field in time-dependent linear dielectric and conductive media, using Choi-Yeon \cite{choi05, choi06} quantization scheme, resorting to a Caldirola-Kanai type Hamiltonian.   Their formal time dependence is realized in terms of parametric 
functions $\eps_l(t)$, which are subject to classical nonstationary oscillator equation, the modulus $|\eps_l(t)|$ 
obeying the Ermakov equation. The relation of these  $\eps_l(t)$  to Choi-Yeon 
\cite{choi05, choi06} parameter functions $\gam_l$, $M_l$ is established. The initial conditions 
(\ref{eps(0)}) under which the eigenstates $|\alf_l;t\ra$ of $\hat A_l(t)$ (and $|n_l;t\ra$ of $\hat A_l^\dg(t) \hat A_l(t)$\,) at $t=0$ coincide with Glauber CS (and photon number states) have been specified. We showed that the medium conductivity and time-dependent permeability act as squeezing and correlating factors. For the quantized 
one mode e.-m. field the first and second moments of photon quadrature operators and the electric and magnetic 
field components in $|\alf_l;t\ra$  are calculated in terms of $|\eps|$. It is shown that they are 
damping in time, and the corresponding variances and covariances minimize Robertson-Schr\"odinger UR.   
The initial photon number states evolve into $|n_l;t\ra$, which are no more eigenstates of the (initial) 
photon number operator. The calculated mean field energy in both $|\alf_l;t\ra$ and $|n_l;t\ra$ exhibits exponential damping factor $\exp(-\Lam(t))$. The dimensionless mean field energy of one mode field in the number states $|n_l;t\ra$ in all times is greater or equal to $n_l$ (the mean energy in initial Fock state), while in $|\alf_l;t\ra$ it can be greater or less than $|\alf_l|^2$, the mean energy in initial Glauber CS. \\
The time evolution of canonical squeezed  states in such media can be treated in a similar way. 
Unlike the Glauber CS, the initially squeezed states of electromagnetic field remain temporally stable, the second moments of the electric and magnetic field components minimizing Robertson-Schr\"odinger UR in all times.

\end{document}